\documentclass{article}
\usepackage{spconf,amsmath,graphicx}
\usepackage{multirow}

\newcommand{\sbf}[1]{\boldsymbol{#1}}

\newcommand{\Cov}{\textrm{Cov}}
\newcommand{\E}{\textrm{E}}

\title{EnLLVM: Ensemble based Nonlinear Bayesian Filtering using Linear Latent Variable Models}
\twoauthors
 {Xiao Lin}
	{University of South Carolina\\
	Department of Computer Science and Engineering\\
	Columbia, SC, USA}
 {Gabriel Terejanu\sthanks{This material is based upon work supported by NIFA under Grand No. 2017-67017-26167}}
	{University of North Carolina at Charlotte\\
	Department of Computer Science\\
	Charlotte, NC, USA}
\begin{document}
%
\maketitle
\begin{abstract}
Real-time nonlinear Bayesian filtering algorithms are overwhelmed by data volume, velocity and increasing complexity of computational models. In this paper, we propose a novel ensemble based nonlinear Bayesian filtering approach which only requires a small number of simulations and can be applied to high-dimensional systems in the presence of intractable likelihood functions. The proposed approach uses linear latent projections to estimate the joint probability distribution between states, parameters, and observables using a mixture of Gaussian components generated by the reconstruction error for each ensemble member. Since it leverages the computational machinery behind linear latent variable models, it can achieve fast implementations without the need to compute high-dimensional sample covariance matrices. The performance of the proposed approach is compared with the performance of ensemble Kalman filter on a high-dimensional Lorenz nonlinear dynamical system.
\end{abstract}
\section{Introduction}
\label{sec:intro}

Bayesian filtering is concerned with obtaining the posterior distribution of quantities of interest (QoI) such as state variables of a dynamical systems as well as unknown parameters in computational models using observational data. Its applicability ranges from data assimilation with applications in weather forecasting~\cite{Isaksen2010} to training recurrent neural networks for computer vision applications~\cite{Gu2017}. For a linear system, under the assumption of Gaussian probability distributions, the problem of estimating the states of the system has an exact closed-form solution given by the Kalman filter~\cite{Kalman1960}. If the probability distributions are non-Gaussian or the system is nonlinear, in general no closed-form solutions are available. Among the most frequently cited approximate nonlinear filters are the extended Kalman filter (EKF)~\cite{Crassidis2004}, the unscented Kalman filter (UKF)~\cite{Julier2004, Merwe2001, Stano2013}, the ensemble Kalman filter (EnKF)~\cite{Evensen2003}, and the particle filter (PF)~\cite{DelMoral1996}.

Both EKF and UKF provide Gaussian approximations to the posterior distributions. In addition, the EKF suffers from poor approximations for high nonlinear systems due to its reliance on model linearization, and the number of samples/sigma points in UKF scales linearly with the number of QoIs. On the other hand EnKF is a reduced rank sampling filter which propagates the uncertainty through the system nonlinearities and updates a relatively small ensemble of samples. Due to its simplicity in both theory and implementation, EnKF has been widely used, especially in meteorology~\cite{Houtekamer2001, Houtekamer2016} where high dimensional data assimilation is performed. Nonetheless, EnKFs have been shown to be sensitive to the violation of the Gaussian assumption~\cite{Lei2010}.

The particle filter uses a Monte Carlo sampling approach by propagating and updating a number of particles without the assumption of Gaussian statistics. Compared with Kalman filter variations, particle filter works with any arbitrary prior distributions and is capable to capture multimodal posterior distributions.  However, particle filter has mostly been applied to low dimensional problems. To cope with high dimensional problems, variations of particle filter have been developed. One class is multiple particle filter~\cite{Djuric2007, Bugallo2014, Ait-El-Fquih2016}. Multiple particle filter uses standard particle filters to update state variables in subspaces. In this way high dimensional filtering is converted to a number of easier lower dimensional problems that require a smaller number of particles. However, problems arise by ignoring when states are coupled in the observation model or when only partial states are observed.    

Often, general formulations that exploit the structure of physics-based models by using internal discrepancies to capture structural errors yield intractable likelihood functions~\cite{OliverJ_CMAME_2015}. In fact, in many cases, as compared with the most common additive observational noise, the distribution of observational noise is unknown and embedded into the nonlinear observational model which yields an intractable likelihood function. These make particle filter unsuitable as it relies on the evaluation of the likelihood function.

We propose a novel Bayesian filtering approach which only requires a small number of samples even in high dimensional systems. The proposed approach uses linear latent projections to approximate the non-Gaussian joint probability distribution between states, parameters, and observables using a mixture of Gaussian components generated by the reconstruction error for each ensemble member. The method is performed without evaluating the likelihood of observations, thus it can be applied to filtering problems in which the likelihood function is intractable.

The rest of paper is organized as follows. The proposed nonlinear filter is detailed in Section~\ref{sec:methodology}. The method is not restricted to a particular type of linear latent models and several options are provided. In Section~\ref{sec:results}, two numerical experiments are performed to assess its performance as compared with EnKF on several different scenarios. Conclusions are provided in Section~\ref{sec:conclusions}.


\section{Methodology}
\label{sec:methodology}

Consider the following general parameterized nonlinear dynamical system perturbed by process noise $\mathbf{w}_k$, measurement noise $\mathbf{v}_k$ and uncertain initial conditions. Given a set of measurements, the goal is to infer the state of the system and parameters to improve model predictions and as a result decision making under uncertainty.
\begin{eqnarray}
\mathbf{x}_{k+1} &=& \mathbf{f}(\mathbf{x}_k, \sbf{\theta},\mathbf{w}_k) \label{ch3_process} \\
\mathbf{d}_k &=& \mathbf{h}(\mathbf{x}_k, \sbf{\theta}, \mathbf{v}_k) \label{ch3_measurement} \\
\mathbf{x}_0 &\sim& p(\mathbf{x}_0)
\end{eqnarray}

Given a set of observation $\mathbf{D}_k = \{\tilde{\mathbf{d}}_i | 1\leq i \leq k \}$, Bayesian filtering is the problem of finding the joint probability density function (pdf) of the states $\mathbf{x}_k$ and parameters $\sbf{\theta}$ conditioned on all the observations up to and including current time $t_k$, $p(\mathbf{x}_k, \sbf{\theta} ~|~ \mathbf{D}_k) $. 

Here, we tackle Bayesian filtering from a different angle. Instead of relying on Eq.~\eqref{ch3_measurement} to calculate the likelihood, we operate on the joint distribution of $\mathbf{x}_{k}$, $\sbf{\theta}$ and $\mathbf{d}_{k}$ directly. For simplicity, in the followings, all the quantities of interest (QoIs) at any time $k$ will be denoted by $\mathbf{q}_{k}$. 
\begin{equation}
\mathbf{q}_{k} = [\mathbf{x}_k, \sbf{\theta}]^{T}
\end{equation}
Furthermore, $\mathbf{q}_{k+1|k}$ is obtained through Eq.~\eqref{ch3_process} after observing $\mathbf{D}_{k}$ but before $\tilde{\mathbf{d}}_{k+1}$. Thus $p(\mathbf{x}_{k+1}$, $\sbf{\theta} ~|~ \mathbf{D}_{k})$ can be written as $p(\mathbf{q}_{k+1|k})$. Similarly, $\mathbf{d}_{k+1|k}$ is obtained by propagating $\mathbf{q}_{k+1|k}$ through Eq.~\eqref{ch3_measurement}.

\begin{figure}[h]
\centering
\includegraphics[width=0.8\linewidth]{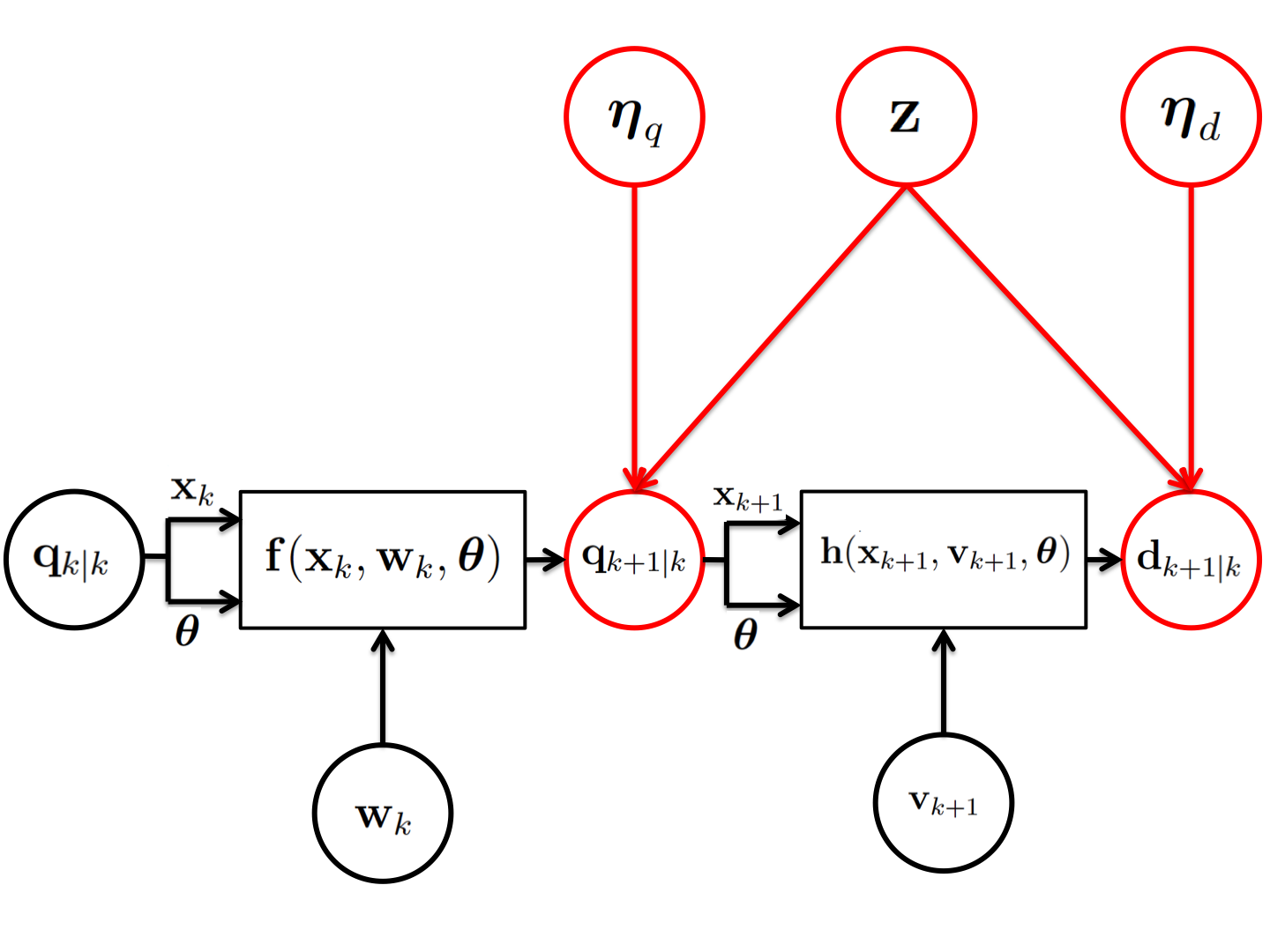} 
\caption{Linear latent variable model construction}
\label{fig:EnLLVM}
\end{figure}

The proposed EnLLVM is a sampling based approach, where $N$ samples from the QoI prior distribution $p(\mathbf{q}_0)$ are propagated through the nonlinear system and updated at every time step. While the propagation of samples is straightforward, the update requires first to obtain a Gaussian mixture approximation of the predictive joint distribution $p(\mathbf{q}_{k+1|k}, \mathbf{d}_{k+1|k})$ based on the reconstruction error of a linear latent variable model (LLVM), see Figure~\ref{fig:EnLLVM}. Second, the posterior distribution of the QoIs is obtained analytically as a Gaussian mixture by conditioning the newly approximated joint pdf with the observational data.

\textbf{EnLLVM: Gaussian mixture approximation}. The linear latent variable model is shown in Eq.~\eqref{Sec2_latentModel}-\eqref{Sec_2_4_dmodel}. $\mathbf{W}_q$ and $\mathbf{W}_d$ are the QoI and observable coefficient matrices, $\mathbf{z}$ is the latent variable, $\sbf{\mu}_q$ and $\sbf{\mu}_d$ are the QoI and observable bias, and $\sbf{\eta}_q$ and $\sbf{\eta}_d$ are the noise variables for the QoI and observable. Assume $\mathbf{q}_{k+1|k}$ and $\mathbf{d}_{k+1|k}$ have a dimensionality of $H_{q}$ and $H_{d}$ respectively, and denote the dimensionality of the joint space as $H = H_{q} + H_{d}$. Let $M$ be the dimensionality of latent variable $\mathbf{z}$, where $M$ is much smaller than $H$. 
\begin{eqnarray}\label{Sec2_latentModel}
\mathbf{q}_{k+1|k} &=& \mathbf{W}_q \mathbf{z} + \sbf{\mu}_q + \sbf{\eta}_q \label{Sec_2_4_qmodel} \\
\mathbf{d}_{k+1|k} &=& \mathbf{W}_d \mathbf{z} + \sbf{\mu}_d + \sbf{\eta}_d \label{Sec_2_4_dmodel} \\
\mathbf{z} &\sim& \mathcal{N}(0,\mathbf{I}_{M \times M}) \nonumber \\
\sbf{\eta}_q &\sim& \mathcal{N}(0,\sbf{\Psi}_q) \nonumber \\
\sbf{\eta}_d &\sim& \mathcal{N}(0,\sbf{\Psi}_d) \nonumber
\end{eqnarray}

The parameters of the LLVM can be obtained via maximum likelihood estimation~\cite{Bishop2006ui} from the samples propagated through the nonlinear system, $\{\mathbf{q}_{k+1|k}^i,\mathbf{d}_{k+1|k}^i\}_{i=1\ldots N}$. We may restrict the structure of the parameters, such that the latent space will preserve certain properties of the original data set. This will lead to different linear latent variable models: (PPCA) probabilistic principal component analysis where $\textrm{blkdiag}(\sbf{\Psi}_q, \sbf{\Psi}_d) = \sigma^2 \mathbf{I}_{H\times H}$, (FA) factor analysis where $\textrm{blkdiag}(\sbf{\Psi}_q, \sbf{\Psi}_d) = \textrm{diag}(\sigma_1^2 \ldots \sigma_H^2)$, and (PCCA) probabilistic canonical correlation analysis where $\sbf{\Psi}_q, \sbf{\Psi}_d$ have off-diagonal elements~\cite{Bach2005wz}.

\begin{figure}[h]
\centering
\includegraphics[width=1.0\linewidth]{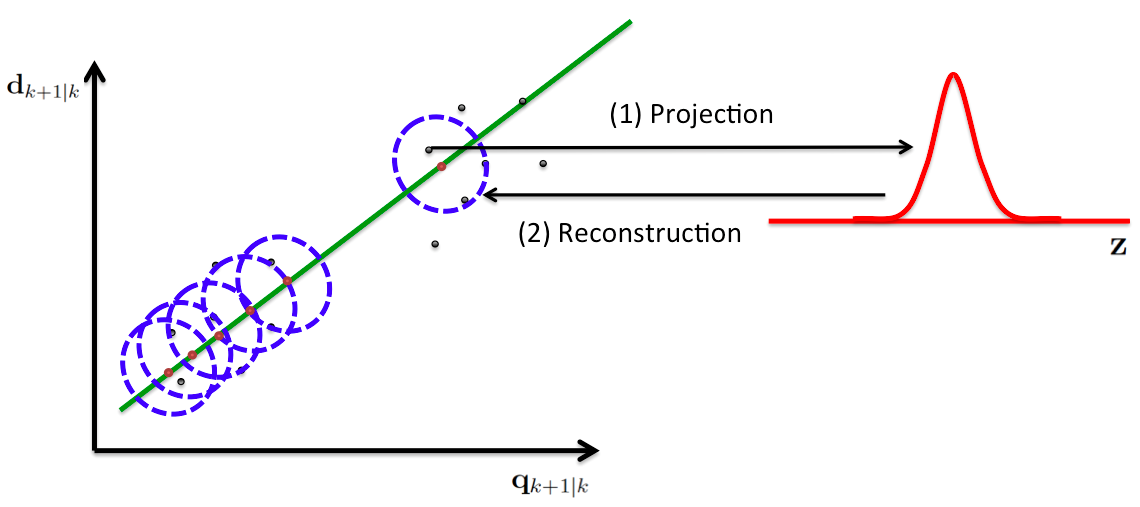} 
\caption{EnLLVM: Gaussian mixture approximation}
\label{fig:GMM}
\end{figure}

Using the above LLVM, the Gaussian mixture approximation of the predicted joint distribution is obtained in the following way. For each sample $\{\mathbf{q}_{k+1|k}^i,\mathbf{d}_{k+1|k}^i\}$ of $p(\mathbf{q}_{k+1|k}, \mathbf{d}_{k+1|k})$, we project it into the latent space where there is a corresponding latent variable $\mathbf{z}_i$ which follows a Gaussian distribution:
\begin{equation}
\mathbf{z}_i \sim  \mathcal{N}( \E[\mathbf{z}|\mathbf{q}_{k+1|k}^i,\mathbf{d}_{k+1|k}^i],  \Cov[\mathbf{z}|\mathbf{q}_{k+1|k}^i,\mathbf{d}_{k+1|k}^i] ) \nonumber
\end{equation}
Projecting the Gaussian latent distribution back we obtain a Gaussian distribution in the original space corresponding to the reconstruction error of the above $i$th sample. Then, the joint pdf $p(\mathbf{q}_{k+1|k}, \mathbf{d}_{k+1|k})$ can be approximated by combining all the Gaussian reconstructions corresponding to the original samples, which results in a Gaussian mixture provided by the EnLLVM, Eq.~\eqref{eq:GMM}, see also Figure~\ref{fig:GMM}.
\begin{eqnarray}
  \hat{p}_{en}(\mathbf{q}_{k+1|k},\mathbf{d}_{k+1|k}) = \frac{1}{N} \sum_{i=1}^N \mathcal{N}(\sbf{\mu}_i, \sbf{\Sigma_i}) \label{eq:GMM} \\
  \sbf{\mu}_i = \mathbf{W} \E[\mathbf{z}|\mathbf{q}_{k+1|k}^i,\mathbf{d}_{k+1|k}^i]+\sbf{\mu} \label{eq:GMM_mu}\\
  \sbf{\Sigma_i} = \mathbf{W} \Cov[\mathbf{z}|\mathbf{q}_{k+1|k}^i,\mathbf{d}_{k+1|k}^i] \mathbf{W}^T + \sbf{\Psi}\label{eq:GMM_cov}
\end{eqnarray}
Here, the LLVM parameters are $\mathbf{W} = [\mathbf{W}_q^T, \mathbf{W}_d^T]^T$, $\sbf{\Psi} = \textrm{blkdiag}(\sbf{\Psi}_q, \sbf{\Psi}_d)$, and the overall Gaussian approximation of LLVM is given by $\sbf{\mu} = [\sbf{\mu}_q^T, \sbf{\mu}_d^T]^T$ and $\sbf{\Sigma} = \mathbf{W}\mathbf{W}^T + \sbf{\Psi}$. The mean and covariance of the Gaussian distribution corresponding to the $i$th latent variable are given by:  
\begin{eqnarray}
\E[\mathbf{z}|\mathbf{q}_{k+1|k}^i,\mathbf{d}_{k+1|k}^i] = \mathbf{W}^T \sbf{\Sigma}^{-1}([\mathbf{q}_{k+1|k}^i,\mathbf{d}_{k+1|k}^i]^T-\sbf{\mu}) \\
\Cov[\mathbf{z}|\mathbf{q}_{k+1|k}^i,\mathbf{d}_{k+1|k}^i] = \mathbf{I}_{M \times M} - \mathbf{W}^T \sbf{\Sigma}^{-1} \mathbf{W}~.
\end{eqnarray}

Computationally, we assume that the $H \gg N \gg M$, where $H$ is the dimensionality of the joint space, $N$ the number of samples, and $M$ the dimensionality of the latent space. Since we can leverage the entire computational machinery of the LLVM, different implementations will yield different computational costs. Nonetheless, expectation maximization (EM) and its variants are preferred. Every iteration of EM will result in $\mathcal{O}(H N M)$, and online EM can process one sample at a time, which is preferred when both $H$ and $N$ are large and samples may require different run times (e.g. different parameter values may increase the stiffness of a system). 

\textit{Lemma}. EnLLVM provides the same estimate for the mean and covariance of the samples as LLVM, while it captures higher order statistics as compared with LLVM.
\begin{eqnarray}
	\E_{en}[\mathbf{q}_{k+1|k},\mathbf{d}_{k+1|k}] &=& \sbf{\mu} \\
	\Cov_{en}[\mathbf{q}_{k+1|k},\mathbf{d}_{k+1|k}] &=& \sbf{\Sigma}
\end{eqnarray}

\textbf{EnLLVM: Bayesian update}. Given the observation data $\tilde{\mathbf{d}}_{k+1}$ and the Gaussian mixture approximation in Eq.\ref{eq:GMM}, the QoI posterior approximation is easily obtained via conditioning of the joint and results in the following Gaussian mixture.
\begin{eqnarray}
  \hat{p}_{en}(\mathbf{q}_{k+1|k+1}) &=&  \sum_{i=1}^N w_{i}\mathcal{N}(\sbf{\mu}_i, \sbf{\Sigma_i}) \label{eq:postGMM} \\
  \sbf{\mu}_i &=& \mathbf{W}_{q} \E[\mathbf{z}_i|\tilde{\mathbf{d}}_{k+1}]+\sbf{\mu}_{q} \\
  \sbf{\Sigma_i} &=& \mathbf{W}_{q} \Cov[\mathbf{z}_i|\tilde{\mathbf{d}}_{k+1}] \mathbf{W}^T_{q} + \sbf{\Psi}_{q}~.
\end{eqnarray} 
Each Gaussian component is assigned with a weight, $w_i$, which is calculated according to the corresponding likelihood under the Gaussian mixture approximation. The mean and covariance matrix of $\mathbf{z}_{i}$ given $\tilde{\mathbf{d}}_{k+1}$ are as follows.
\footnotesize
\begin{eqnarray}
\E[\mathbf{z}_i|\tilde{\mathbf{d}}_{k+1}] &=& \Cov[\mathbf{z}_i|\tilde{\mathbf{d}}_{k+1}] (\mathbf{W}^T \sbf{\Psi}_d^{-1}(\tilde{\mathbf{d}}_{k+1} -\sbf{\mu}_d) \label{eq_34} \\
&&+ \Cov[\mathbf{z}|\mathbf{q}_{k+1|k}^i,\mathbf{d}_{k+1|k}^i]^{-1}\E[\mathbf{z}|\mathbf{q}_{k+1|k}^i,\mathbf{d}_{k+1|k}^i]) \nonumber \\
\Cov[\mathbf{z}_i|\tilde{\mathbf{d}}_{k+1}] &=& (\Cov[\mathbf{z}|\mathbf{q}_{k+1|k}^i,\mathbf{d}_{k+1|k}^i]^{-1}  + \mathbf{W}_d^T \sbf{\Psi}_d^{-1} \mathbf{W}_d)^{-1} \nonumber
\end{eqnarray}
\normalsize

\textbf{EnLLVM: Noise inflation}. To avoid component degeneracy and accommodate for potential model error, noise inflation is performed. We introduce a new hyperparameter $\alpha$ in the distribution of the data noise $\sbf{\eta}_d$:
\begin{equation}
\sbf{\eta}_d \sim \mathcal{N}(0,\alpha\sbf{\Psi}_d) 
\end{equation}
Here, $\alpha$ is obtained via maximum likelihood:
\begin{eqnarray}
 \alpha^* &=& \arg_\alpha\max \mathcal{N}(\tilde{\mathbf{d}}_{k+1}; \sbf{\mu}_d, \mathbf{W}_d\mathbf{W}_d^T + {\alpha}\sbf{\Psi}_d) \\
 &=& \frac{1}{H}\textrm{tr}(\sbf{\Psi}_d^E \sbf{\Psi}_d^{-1})
\end{eqnarray}

If $\alpha^{*} >1$, the noise in Eq.~\eqref{Sec_2_4_dmodel} will be inflated, which will lead to larger uncertainty in the latent variable. This uncertainty will eventually be reflected in the uncertainty of QoIs through Eq.~\eqref{Sec_2_4_qmodel}. This means the model error will be absorbed in the uncertainty of QoIs, which provides a way to investigate model error and develop models. 


\section{Numerical Results}
\label{sec:results}

\subsection{Example 1 -  Lorenz 63}
\label{ex1_Bimodal}

The Lorenz 63 is a three dimensional system and it is used here to show that EnLLVM can capture non-Gaussian distributions~\cite{terejanu2011adaptive, Dovera2011}.
\begin{align}
dx_{1}/dt &= -cx_1 + cx_2 \\
dx_{2}/dt &= -x_1x_3 + rx_1 - y_2 \\
dx_{3}/dt &= x_1x_2 - bx_3 
\end{align}
Here $c = 10$, $b = \frac{8}{3}$ and $r = 28$. The discrete measurement model is given in Eq.~\eqref{observationL63}
\small
\begin{equation}\label{observationL63}
d_{k} = \sqrt{x_1(t_k)^2 + x_2(t_k)^2 + x_3(t_k)^2} + \upsilon_{k}, ~\upsilon_{k}  \sim \mathcal{N}(0, 1) 
\end{equation}
\normalsize
The state prior distribution is set to: 
\footnotesize
\begin{equation}
p(x(t_{0})) \sim 0.5 \mathcal{N}([-0.2, -0.2, 8]^{T}, \sqrt{0.35}\textbf{I}_{3}) + 0.5 \mathcal{N}([0.2, 0.2, 8]^{T}, \sqrt{0.35}\textbf{I}_{3})  \label{initial:L63}
\end{equation}
\normalsize
The total simulation time is $4$sec and the system is discretized with a time step $\Delta t = 0.1$sec. Synthetic measurements are generated by randomly selected initial states from Eq.~\eqref{initial:L63} and propagated through Eq.~\eqref{observationL63}. States are updated every $4$ time steps. 

\begin{figure}[h]
 \vspace{-0.15in}
  \begin{center}
    \includegraphics[width=1.0\linewidth]{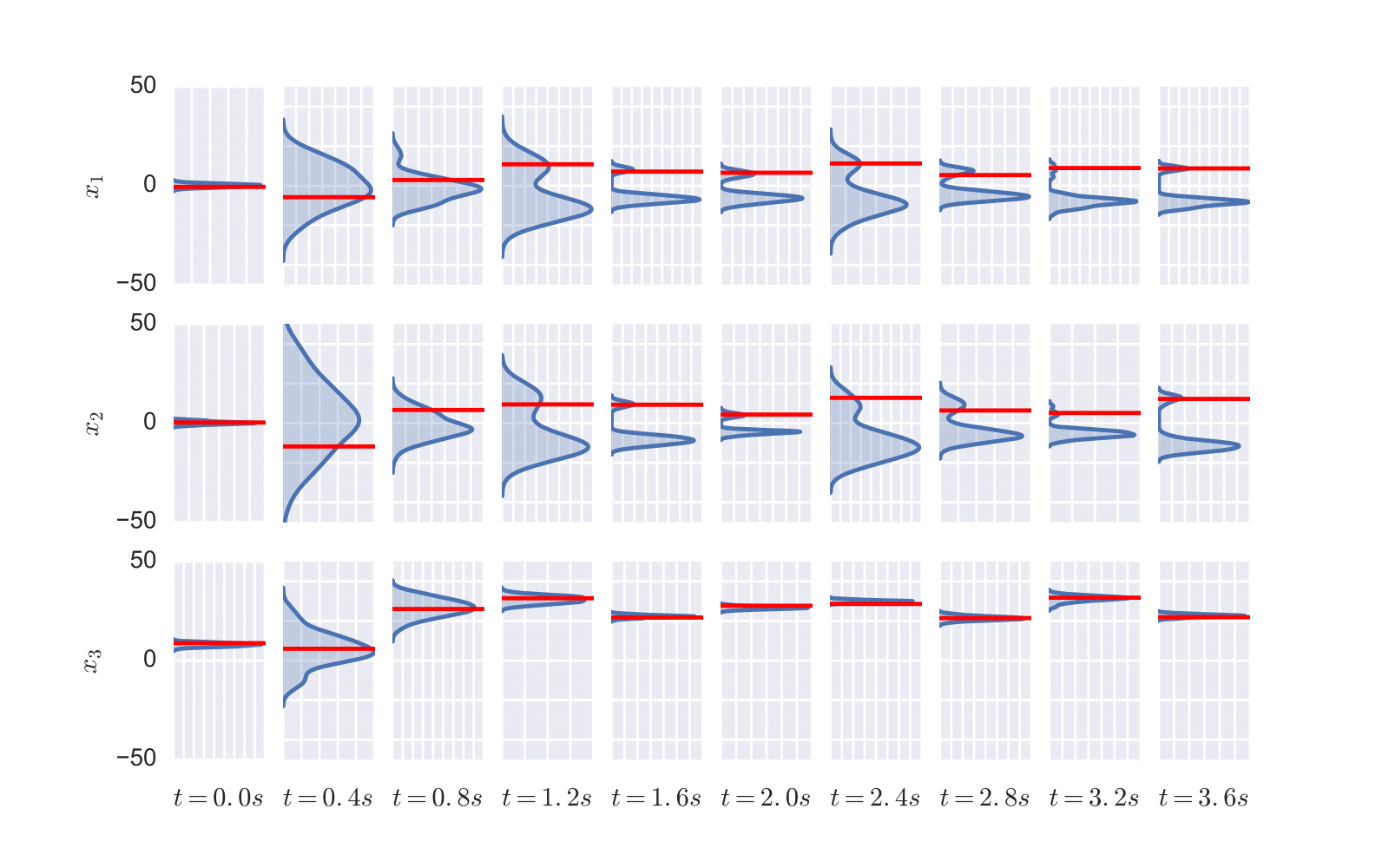}
  \end{center}
  \vspace{-0.3in}
  \caption{Posterior distribution of states in Lorenz 63 system after each update.}
  \label{fig:sec_3_1_Lorenz63}
\end{figure}

The joint space of states and observations have a dimensionality of four, and in this simulation, we use two components for PPCA and $30$ samples for Bayesian update. Fig.~\ref{fig:sec_3_1_Lorenz63} shows the posterior distribution of states after each update. The horizontal lines indicate the true states. As we can see, EnPPCA is capable to capture the true states and the bimodal distribution of the states.
\subsection{Example 2 - Lorenz 96}
\label{ex2_enppca_enkf}

In this example, EnPPCA and EnKF will be applied to track the states of Lorenz 96, a $40$ dimensional nonlinear system and commonly used in benchmark studies. 
\small
\begin{equation} \label{Sec_3_Lorenz}
dx_j(t)/dt = (x_{j+1}-x_{j-2}) x_{j-1} - x_j + 8, ~j=1\ldots 40~, 
\end{equation}
\normalsize
where $x_0 = x_{40}, x_{-1} = x_{39},$ and $x_{41} = x_1$.  The system is discretized with a time step of $\Delta t = 0.001$sec. The prior is given by $\mathcal{N}(0,1)$ and the observation model provides an incomplete observation of the state of the system at every $\Delta t = 0.1$sec. Besides the linear measurement model, two different nonlinear measurement models are also used. Another important scenario is the presence of model error. As previously discussed, EnPPCA has the property to absorb potential model error and to reflect it in the posterior distribution of the QoI. Here, four different models are used to generate the observational data and they only differ in the constant forcing. 

The joint space between the QoIs (states of the system) and the observable has dimensionality of $60$.  For this study, $5$ components are used for EnPPCA, which means the dimensionality of the latent space is $5$. Here, we use $30$ samples to track the states of the system. Root mean square error (RMSE) is used as the metric to measure the predictability of EnPPCA and EnKF. RMSE is calculated between the mean of posterior samples and the true state value. Table.~\ref{my-label} shows the RMSE statistics of $100$ trials. If no model error exists, the performance of EnKF is comparable with or better than EnPPCA in average RMSE for all three measurement models, otherwise the performance of EnLLVM is statistically significant better than EnKF as the model error increases. 
\begin{align}
\begin{cases}
&\textrm{Data generation process:} \nonumber\\
&M1: ~ dx_j(t)/dt =~(x_{j+1}-x_{j-2}) x_{j-1} - x_j + 9 \nonumber \\
&M2: ~ dx_j(t)/dt =~ (x_{j+1}-x_{j-2}) x_{j-1} - x_j + 10 \nonumber \\
&M3: ~ dx_j(t)/dt =~ (x_{j+1}-x_{j-2}) x_{j-1} - x_j + 11 \nonumber \\
&M4: ~ dx_j(t)/dt =~ (x_{j+1}-x_{j-2}) x_{j-1} - x_j + 12 \nonumber \\
&\textrm{Linear measurement model:} \nonumber\\
&d_j(t) = x_{2j-1}(t) + v_j(t), \quad v_j(t) \sim \mathcal{N}(0,I_{20}) \nonumber\\
&\textrm{Nonlinear measurement models:} \nonumber \\
&(I): ~ d_j(t) = x_{2j-1}(t)x_{2j}(t) + v_j(t), \quad v_j(t) \sim \mathcal{N}(0,I_{20}) \nonumber \\
&(II): ~ d_j(t) = x_{2j-1}(t)^2 + v_j(t), \quad v_j(t) \sim \mathcal{N} (0,I_{20}) \nonumber
\end{cases}
\end{align}
\begin{table}[]
\centering
\caption{EnKF vs EnPPCA: RMSE statistics of 100 trials}
\label{my-label}
\begin{tabular}{lllll}
\multirow{2}{*}{\textbf{Model Error}} & \multicolumn{2}{c}{\textbf{EnKF}}                                   & \multicolumn{2}{c}{\textbf{EnPPCA}}                                 \\
                                      & \multicolumn{1}{c}{\textbf{Mean}} & \multicolumn{1}{c}{\textbf{STD}} & \multicolumn{1}{c}{\textbf{Mean}} & \multicolumn{1}{c}{\textbf{STD}} \\
                                      \hline \hline
\multicolumn{5}{l}{\textbf{Linear Measurement Model}}                                                                                                                             \\
No err.                               & \textbf{2.71}                     & 0.61                   & 2.92                              & \textbf{0.23}                            \\
M1: 9                                 & 3.71                              & 0.63                            & \textbf{2.99}                     & \textbf{0.23}                   \\
M2: 10                                & 4.63                              & 0.71                            & \textbf{3.07}                     & \textbf{0.22}                   \\
M3: 11                                & 5.31                              & 0.81                            & \textbf{3.18}                     & \textbf{0.21}                   \\
M4: 12                                & 5.9                               & 0.84                            & \textbf{3.27}                     & \textbf{0.20}                   \\
\hline
\multicolumn{5}{l}{\textbf{Nonlinear Measurement Model (I)}}                                                                                                                      \\
No err.                               & \textbf{1.87}                     & \textbf{0.53}                   & 2.85                              & 0.26                            \\
M1: 9                                 & 3.3                               & 0.63                            & \textbf{2.91}                     & \textbf{0.22}                   \\
M2: 10                                & 4.37                              & 0.47                            & \textbf{3.02}                     & \textbf{0.21}                   \\
M3: 11                                & 5.01                              & 0.44                            & \textbf{3.17}                     & \textbf{0.17}                   \\
M4: 12                                & 5.51                               & 0.49
 & \textbf{3.32}                     & \textbf{0.17}                   \\
 \hline
\multicolumn{5}{l}{\textbf{Nonlinear Measurement Model (II)}}                                                                                                                     \\
No err.                               & 2.87                              & 0.76                            & \textbf{2.84}                              & \textbf{0.24}                            \\
M1: 9                                 & 3.78                              & 0.75                            & \textbf{2.94}                     & \textbf{0.22}                   \\
M2: 10                                & 4.78                              & 0.56                            & \textbf{3.03}                     & \textbf{0.20}                   \\
M3: 11                                & 5.43                              & 0.53                            & \textbf{3.16}                     & \textbf{0.19}                   \\
M4: 12                                & 6.1                               & 0.55                            & \textbf{3.37}                     & \textbf{0.17}       \\
\hline            
\end{tabular}
\end{table}


\section{Conclusions}
\label{sec:conclusions}

A general and fast computational framework is developed for nonlinear filtering of high dimensional nonlinear dynamical systems. This is achieved by developing a robust probabilistic model to approximate the Bayesian inference problem and obtain samples from high dimensional posterior distributions. In addition, the proposed approach addresses one of the main challenges in uncertainty quantification, namely dealing with modeling errors and their effect on inference and prediction.




\vfill\pagebreak


\bibliographystyle{IEEEbib}
\bibliography{paper}

\end{document}